# Cyber-Physical Interdependent Restoration Scheduling for Active Distribution Network Via Ad Hoc Wireless Communication

Chongyu Wang, Mingyu Yan, Kaiyuan Pang, Fushuan Wen, *Fellow, IEEE*, and Fei Teng, *Senior Member, IEEE*

*Abstract*—This paper proposes a post-disaster cyber-physical interdependent restoration scheduling (CPIRS) framework for active distribution networks (ADN) where the simultaneous damages on cyber and physical networks are considered. The ad hoc wireless device-to-device (D2D) communication is leveraged, for the first time, to establish cyber networks instantly after the disaster to support ADN restoration. The repair and operation crew dispatching, the remote-controlled network reconfiguration and the system operation with DERs can be effectively coordinated under the cyber-physical interactions. The uncertain outputs of renewable energy resources (RESs) are represented by budget-constrained polyhedral uncertainty sets. Through implementing linearization techniques on disjunctive expressions, a monolithic mixed-integer linear programming (MILP) based two-stage robust optimization model is formulated and subsequently solved by a customized column-and-constraint generation (C&CG) algorithm. Numerical results on the IEEE 123-node distribution system demonstrate the effectiveness and superiorities of the proposed CPIRS method for ADN.

## NOMENCLATURE

### Indices and sets

| | |
|---|---|
| $rc$, $oc$ | Indices for repair and operating crews. |
| $m, m', n / \Omega^F$ | Indices / set for faulted components. |
| $p, p', q / \Omega^{MS}$ | Indices / set for manual switches (MSs). |
| $s / \Omega^{RCS}$ | Index / set for remote-controlled switches (RCSs). |
| $t / T$ | Index / set for time slots. |
| $i, j / N$ | Indices / set for nodes. |
| $i', j' / \mathcal{N}^{NC}$ | Indices / set for node cells (NCs). |
| $nc / NC$ | Index / set for NCs. |
| $ij, l / L$ | Indices / set for lines in original networks. |
| $i'j' / \mathcal{E}^{NC}$ | Index / set for lines in simplified networks. |
| $\Omega_{nc}^{NC,F}$ | Set for faulted components inside $nc$-th NC. |
| $\Omega_p^{NC,MS}$ | Set for NCs connected by $p$-th MS. |
| $\Omega_s^{NC,RCS}$ | Set for NCs connected by $s$-th RCS. |
| $N^S / N^{GT} / N^{RES}$ | Index for substation / gas turbine (GT) / RES nodes. |
| $U_i^N / D_i^N$ | Set for upstream /downstream nodes of $i$. |
| $U_{i'}^{NC} / D_{i'}^{NC}$ | Set for upstream / downstream NCs of $i'$. |
| $c, c'$ | Indices for cyber routers. |
| $k$ | Index for cyber links. |
| $L_{c,k}^C$ | Set for cyber routers contained in $k$-th cyber link of $c$-th router. |
| $C^S / C^{RCS} / C^{GT} / C^{RES}$ | Sets for cyber routers of substations / RCSs / GTs / RESs. |
| $C^{Add}$ | Set for added relay cyber routers. |

### Parameters

| | |
|---|---|
| $c_i^{Penalty}$ | Penalty cost of $i$-th bus. |
| $\Delta T$ | Time interval of scheduling. |
| $T_{m,n,rc}^{TR,RC} / T_{p,q,oc}^{TR,OC}$ | Travel duration of $rc$-th RC from faulted $m$-th component to $n$-th component / $oc$-th OC from $p$-th MS to $q$-th MS. |
| $T_{p,oc}^{MSO}$ | Manual operation duration of $oc$-th OC at $p$-th sectionalizing switch. |
| $T_s^{RCSO}$ | Remote operation duration of $s$-th RCS. |
| $P_{i,t}^{LD}$ | Maximum load demand of $i$-th node at time $t$. |
| $\underline{V_i}, \overline{V_i}$ | Voltage security margin of $i$-th bus at time $t$. |
| $\overline{P_l^L}, \overline{Q_l^L}$ | Maximum allowable capacity of active, reactive power of line $ij$. |
| $\overline{P_{i,t}^{S/GT}}$ | Active capacity of the $i$-th substation / GT at time $t$. |
| $\overline{Q_{i,t}^S} / \overline{Q_{i,t}^{GT}} / \overline{Q_{i,t}^{RES}}$ | Reactive capacity of the $i$-th substation / GT / RES at time $t$. |
| $R_i^+ / R_i^-$ | Up / down ramping rate of $i$-th GT. |
| $\rho_i^{GT}$ | Reserve factor of $i$-th GT. |
| $r_{ij}, x_{ij}$ | Resistance and reactance of line $ij$. |
| $\gamma_i^{GT} / \gamma_i^{RES}$ | FRR of the $i$-th GT / RES. |
| $|NC|$ | Total number of NCs. |
| $T^{MAX}$ | Sufficient large time not less than time horizon of the entire restoration procedure. |
| $\hat{P}_{i,t}^{RES}$ | Forecasted active power output of the $i$-th RES at time $t$. |
| $\omega_i$ | Maximum prediction error (MPE) of the power output of $i$-th RES. |
| $B_i^{RES}$ | Budget of the uncertainty (BoU) of $i$-th RES. |

Corresponding author: Dr Fei Teng.
C. Wang and K. Pang are with the College of Electrical Engineering, Zhejiang University, Hangzhou 310027, China. (e-mail: chongyu.wang@outlook.com; Kaiyuan.Pang@zju.edu.cn).
M. Yan and F. Teng are with the Department of Electrical and Electronic Engineering, Imperial College London, London SW7 2AZ, U.K. (e-mail: mingyu.yan@imperial.ac.uk; f.teng@imperial.ac.uk).
F. Wen is with the Hainan Institute, Zhejiang University, Sanya 572000, China, and also with the College of Electrical Engineering, Zhejiang University, Hangzhou 310027, China. (e-mail: fushuan.wen@gmail.com)

| Symbol | Description |
|---|---|
| $n_c^C$ | Total number of cyber links of $c$-th router. |
| $\overline{P}_c^{Router}$ | Power consumption of $c$-th router. |
| $T_c^{UPS,C}$ | Power supply time of the UPS of $c$-th router. |

*Binary Variables*

| Symbol | Description |
|---|---|
| $x_{m,n,rc}^{RC}$ | 1 if $rc$-th RC is assigned to repair faulted component $n$ after $m$, 0 otherwise. |
| $x_{p,q,oc}^{OC}$ | 1 if $oc$-th OC is assigned to operate $q$-th MS after the work of $p$-th MS, 0 otherwise. |
| $u_{n,t}^L$ | 1 if faulted component $n$ is repaired before time $t$, 0 otherwise. |
| $u_{nc,t}^{NC}$ | 1 if all potential faults in $nc$-th NC clear before time $t$, 0 otherwise. |
| $w_{l,t}^{MS} / w_{l,t}^{RCS}$ | 1 if the switch on line $l$ is closed manually/ remotely before time $t$, 0 otherwise. |
| $w_{l,t}$ | 1 if line $l$ is closed at time $t$, 0 otherwise. |
| $u_{c,t}^C$ | 1 if $c$-th router is available at time $t$, 0 otherwise. |
| $u_{c,k,t}^{C,Link}$ | 1 if $k$-th cyber link of $c$-th router is available at time $t$, 0 otherwise. |
| $u_{c,t}^{UPS,C}$ | 1 if the UPS of $c$-th router is available at time $t$, 0 otherwise. |
| $\xi_{i',t}$ | 1 if aggregated node (node cell) $i'$ is a root node at time $t$, 0 otherwise. |

*Functions and Other Variables*

| Symbol | Description |
|---|---|
| $AT_{n,rc}^{RC} / AT_{q,oc}^{OC}$ | Arrival time of $rc$-th RC / $oc$-th OC at the faulted component $n$ / $q$-th MS. |
| $T_{m,rc}^{RP}$ | Repair work duration of $rc$-th RC at faulted component $m$. |
| $t_n^{E,RP}$ | Repair ending time of faulted component $n$. |
| $T_{p,oc}^{OC,S}$ | Time when $oc$-th OC arrives and can safely perform the operation of $p$-th MS. |
| $T_p^{NCR,MS} / T_s^{NCR,RCS}$ | Final fault clearing time for all NCs connected with $p$-th MS / $s$-th RCS. |
| $T_{nc}^{NCR}$ | Fault clearing time for $nc$-th NC. |
| $t_q^{E,MSO}$ | Operating ending time of $q$-th MS. |
| $V_{i,t}$ | Voltage value of node $i$ at time $t$. |
| $P_{i,t}^{Shed}$ | Load Shedding on bus $i$ at time $t$. |
| $P_{i,t}^S / P_{i,t}^{GT} / P_{i,t}^{RES}$ | Active power at node $i$ generated by substation / GT / RES at time $t$. |
| $Q_{i,t}^S / Q_{i,t}^{GT} / Q_{i,t}^{RES}$ | Reactive power at node $i$ generated by substation / GT / RES at time $t$. |
| $\overline{P}_{i,t}^{RES}$ | Maximum available active power output of RES $i$ at time $t$. |
| $P_{ij,t}^L / Q_{ij,t}^L$ | Active / reactive power on line $ij$ at time $t$. |
| $P_{i,t}^G / Q_{i,t}^G$ | Active / reactive power injected to the network at bus $i$ at time $t$. |
| $F_{i'j',t}$ | Single-commodity flow on line $i'j'$ at time $t$. |
| $\sigma_{i,t}^{+,RES} / \sigma_{i,t}^{-,RES}$ | Upward/ downward uncertainty of available power output of RES $i$ at time $t$. |

## I. INTRODUCTION

To prevent blackout is one of the most significant motivations for the evolution of power systems. Over the past few years, high-impact low-probability (HILP) events led by natural disasters occurred frequently all across the world [1], such as tornado Blyth (South Australia, Australia) in 2016, hurricane Harvey (Texas, USA) in 2017, typhoon Hagupit (Zhejiang, China) in 2021. After these extreme events, a substantial number of consumers lost power supplies, accompanied by huge economic losses [2]. Power distribution systems (PDS), directly oriented to end-users, are expected to be restored rapidly, which has attracted widespread attention from both academic and industrial communities.

Toward the ambitious vision of building a resilient PDS, utilities implement the fault location, isolation and service restoration (FLISR) [3] step by step after the event. Under HILP events, a considerable amount of infrastructures in PDS, such as poles and feeder lines, may be destroyed, and blackouts would occur thereupon due to the electrical disconnection between loads and the upstream power grid. As a result, the decision-making of the resilience-oriented restoration scheduling is more complicated than the reliability-oriented one, which only focuses on "N-1/N-2" scenarios [4].

An effective service restoration scheme needs to be devised to improve post-disaster system performance by coordinating the system operation while dispatching skilled maintenance crews to clear faults and perform topological changes. With the rapid development of distributed energy resources (DERs), such as gas turbines (GTs) in combined heat and power plants and renewable energy resources (RESs) represented by wind power and photovoltaics, PDS is gradually transformed into active distribution networks (ADN). There has been a significant amount of research regarding efficient restoration scheduling schemes for the physical system of PDS or ADN. A two-stage post-disaster restoration scheduling strategy is developed in [5], where repair tasks of dispatched crews are clustered and determined in the first stage, while the dispatching of crews and operation of DERs are coordinated in the second stage. In [6], the joint dispatching of repair crews and mobile power sources is formulated as a mixed-integer linear programming (MILP). A variable time-step (VTS) restoration model is proposed in [7] to reduce the imposed constraints when implemented in PDS. In [8], the dispatching of repair crews and mobile battery-carried vehicles are coordinated in networked microgrids with soft-open points. In [9], robust restoration schemes are devised for ADN by considering uncertain RESs' outputs.

However, cyber networks are always assumed to be intact after the disaster in the above-mentioned publications. With the rapid growth of cyber networks to support physical system, the cyber and physical lays in power systems are inextricably linked and deeply coupled [10]. A damage in cyber system may lead to a single fault or even cascading failures of physical components. Both the cyber and physical systems may be simultaneously damaged under extreme events, such as the Atlantic hurricane season in 2017 [11], where 50% of wireless communications were lost in Texas due to hurricane Harvey and approximately 95.6% of base stations out of service in Puerto Rico's due to Hurricane Maria. This highlights that the availability of cyber systems should be explicitly evaluated prior to developing restoration scheduling schemes for power systems. Toward this regard, the impact of cyber-physical interactions on system resilience under extreme natural disasters is investigated in [12]. After the communication is interrupted, the dispatch centre has to send power adjustment commands through personal calls, resulting in a significant delay in power regulation.

To tackle this challenge, some recent research starts to investigate the optimal post-event scheduling schemes for the restoration of cyber-physical PDS. In [13], an executable restoration scheduling model for cyber-physical PDS utilizing wireless communication networks is developed. A discrete event simulator is used to quantitatively analyze the cyber-physical interdependencies. In [14], emergency communication vehicles are utilized to set up ad hoc communication links for distribution automation (DA). However, the above methods are devised for PDS instead of ADN, that is, DERs are not considered as potential means to boost the restoration process. Networked microgrids constructed by DERs have been demonstrated to be beneficial for enhancing the ADN's resilience [15], but the sequential restoration scheduling strategies considering cyber-physical coupling still need further exploration.

Communication schemes for the smart grid can be divided into two categories: wired communication, such as power line carriers and optic fiber, and wireless communication [16]. Since the topology of the wired communication network is pre-determined, the reliability of wired communications basically depends on the availability of each line being laid (that is, each feeder line for power line carriers, and each fiber line for optic fiber). For each line in wired communications, aging and attack by external forces are inevitable. In addition, because the end-users of medium voltage ADN are numerous and scattered, the costs of laying optic fiber could be significant [17]. The different levels of development in urban and rural areas [18] also constrain the full construction of high-cost infrastructure. As an alternative, wireless communication with cost-effectiveness, convenient deployment and flexible network construction approaches gains more attention, and is considered to be one of the most promising communication solutions for ADN with frequent end-user extensions recently [18].

The fast development of wireless communications [20] such as ZigBee, cellular networks, private long-term evolution (LTE), and private 5G [21] provides alternative solutions for industrial utilities. For instance, the wireless private network has been used in China to solve "the last mile" communication tasks such as the DA and load control [22]. Recently, 5G wireless communications are implemented in the power distribution system in the county where the laying of fibers is inconvenient [23]. However, the role and value of wireless communications to support system restoration has not been well understood. As an emerging technique, in particular, the device-to-device (D2D) communication enabling the direct data exchange between end-users provides a flexible networking method by which ad hoc communications for remote network reconfigurations and microgrid coordination can be established rapidly. Considering conspicuous cyber-physical interactions, D2D communication has promising application prospects in realizing ADN restoration with instant available post-disaster cyber systems.

To sum up, the key limitation of the exiting research regarding sequential ADN restoration scheduling is the lack of consideration of the cyber system availability and the collaborative restoration of cyber and physical networks. To fill this gap, this paper proposes a novel D2D-assisted cyber-physical interdependent restoration scheduling (CPIRS) framework for ADN. The main contributions are summarized as follows.

1) This paper proposes the *CPIRS method* for e*fficient and executable restoration* of ADN that explicitly considers c*yber-physical interdependencies*. The ad hoc wireless D2D communication is leveraged to assist network reconfiguration and form stably operating microgrids efficiently.

2) To obtain the executable scheduling under extreme scenarios, the uncertain outputs of renewable energy sources (RESs) are explicitly considered and modelled by polyhedral uncertain sets with given budgets of uncertainties.

3) Task clustering and linearization of disjunctive and non-linear expressions are conducted to reformulate the optimization problem as the monolithic MILP-based two-stage robust optimization model, which can be efficiently solved by a customized column-and-constraint generation (C&CG) algorithm.

The remainder of this paper is organized as follows. The problem is described in Section II. Mathematical formulations and their modifications are introduced in detail in Section III and Section IV, respectively. Case studies are presented in Section V. Finally, Section VI concludes this paper.

## II. Description of Cyber-Physical ADN Restoration

The cyber-physical ADN restoration begins right after the extreme event. To focus on the severe damages, both the three-phase grounding faults of feeders on the physical side and the paralysis of conventional communication systems (except for the ad hoc wireless communications) on the cyber side are considered. The specific restoration service involves the repair of faulted components and manual/remote network reconfigurations. The key aspects of the CPIRS will be introduced hereinafter in this section.

### A. Service Restoration of Physical ADN

As illustrated in Fig. 1, when the ADN suffers a major disturbance caused by a HILP event at $t_0$, it will remain in degraded operation for a period of time, during which the fault diagnosis [28], fault isolation [29] will be implemented to prepare for the service restoration. Note that the system performance measures the ADN's ability to quickly restore the power supplies for customers when facing extreme events.

From $t_1$ to $t_2$, the ADN is expected to restore energy supplies to customers as soon as possible. Eventually, the ADN will recover the original system performance level. The resilience metric quantifying the performance of ADN during the whole event can be given by $\int_{t_0}^{t_2}\left[\overline{F}(t)-\underline{F}(t)\right]$, presented as the shaded parts. This paper focuses on the system performance enhancement during the service restoration stage, shown as the red arrows, so the resilience metric turns $\int_{t_1}^{t_2}\left[\overline{F}(t)-\underline{F}(t)\right]$ accordingly.

Power loads, especially critical loads, are expected to be picked up as quickly as possible after the disaster, rather than maintaining a high-level average power quality right after the extreme event. Therefore, the expected energy not supplied (EENS) is selected in this paper as a quantitative criterion for evaluating the system performance in this paper, instead of utilizing the criteria evaluating the average resilience in the ADN, such as the system average interruption frequency index and the system average interruption duration index.

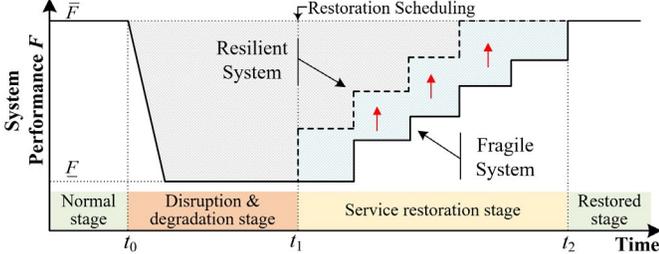

Fig. 1 System performance changes under different restoration strategies.

### B. Cyber-Physical Interdependencies in ADN Restoration

With the ever-widespread application of intelligent electronic devices (IEDs), the interdependencies of cyber systems and physical systems are increasing. Commands for operating RCSs, received by feeder terminal units (FTUs), and voltage/frequency regulations during microgrid expansions or combinations, received by DERs, need to be sent by the control centre (the master station) of ADN, while device status and response signals of the relevant equipment have to be uploaded. The aforementioned information exchanges depend on the availability of cyber system. Without the support from cyber networks, the power node may become unobservable (i.e., the relevant measurement data cannot be shared) and/or uncontrollable (i.e., control signals cannot be reached or sent out). On the other hand, the normal operations of cyber routers in cyber systems also require the energy supply from the physical system.

### C. Post-Disaster Ad Hoc Wireless Communication Network

The communication network may be partially unavailable or even completely destroyed after the disaster, resulting in extensive research on rapidly deployed networks (RDNs) [30]. Establishing an ad hoc communication network to support related functions as soon as possible, rather than waiting to fully restore the original communication system, is the primary task of RDNs. As a potential pattern of RDNs, the emerging wireless device-to-device (D2D) communications [31] start to catch the attention of power distribution systems. For example, in [32], a post-disaster controller-to-controller (C2C) communication-assisted decentralized control strategy is proposed for ADN.

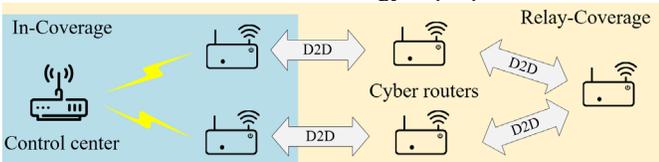

Fig. 2 Ad hoc D2D communications among the control centre and routers.

In conventional cellular networks, direct communication between end-users is not permitted because the uplink and downlink are separate. As illustrated in Fig. 2, D2D communications activate the multi-hop network construction among user terminals [33], where the relay communication no longer depends on the base station, thereby facilitating the post-disaster restoration. The currently used communication system that supports DA involves the application layer, the communication layer and the interface layer [34]. In general, load nodes in the power distribution system do not have adequate and special computing power resources. Therefore, the layered communication structures (i.e., the control centre and the routers) are used in almost all the state-of-the-art communication frameworks in ADN [35]. The master station of the ADN with computing power can be considered as the control centre and as the application layer. The correct execution of the DA requires the bidirectional information exchange between the control centre and each intelligent electronic device (IED), such as the FTU controlling RCSs. All wireless communication routes between the cyber routers at IEDs and the cyber router at the control centre constitute the communication layer. Finally, the interface layer consists of all IEDs enabling actions of RCSs. In terms of the network reconfiguration under post-disaster scenarios, end-to-end cyber interconnections between IEDs controlling each RCS could be built via direct short-range wireless communication empowered by D2D, and thereby, communications within the ADN can be connected up through the ad hoc cyber network [24]. It is worth noting that the ad hoc wireless communication supported by D2D only facilitates flexible networking constructions between terminal IEDs, but will not change the communication framework of the DA under normal operation.

Given the wireless communication quality criteria, such as signal to interference & noise ratio (SINR) and reference signal received quality (RSRQ) [36], an ad hoc cyber network over the studied ADN can be generated to obtain available cyber links between the control centre and each cyber router. Since the focus of this paper is to implement an ad hoc wireless communication network for emergency DA after the disaster, the basic availability of each cyber link of the established ad hoc cyber system is the primary and crucial consideration. Hence, to simplify the modelling, the maximum communication radius is utilized to represent the communication range of a wireless cyber router. Furthermore, because the maximum communication radius varies with the specific cyber devices, the parameter setting should refer to the wireless cyber routers deployed in the studied ADN.

## III. MATHEMATICAL MODEL FORMULATION

The framework of the proposed CPRIS is shown in Fig. 3, which constitutes the crew dispatching model, the physical system operation model, the cyber system availability model and the cyber-physical interdependency model. The details of each component will be discussed in the following subsections.

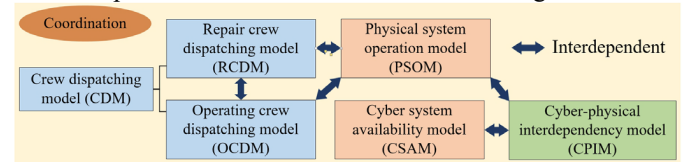

Fig. 3 Framework of the CPIRS for ADN.

### A. Crew Dispatching Model (CDM)

After the ADN was destroyed by an extreme event, repair crews (RCs) and operating crews (OCs) will be dispatched in a unidirectional acyclic routing sequence (shown in Fig. 4) to repair all the faulted components and operate significant manual switches (MSs), respectively. As such, the proposed crew dispatching model contains the repair crew dispatching model (RCDM) and the operating crew dispatching model (OCDM). By nature, this dispatching problem is one kind of well-known vehicle routing problem (VRP) [37].

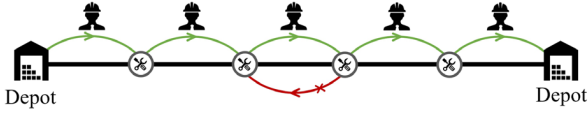

Fig. 4 The unidirectional acyclic routing sequence of RC/OC.

*1) Repair crew dispatching model (RCDM)*

The routing sequence of each dispatched RC is formulated by constraints (1)-(4), where $rc \in RC$, $m, m', n \in \Omega^F$:

$$\sum_{m \in \Omega^F} x^{RC}_{dp^{RC}_{rc}, m, rc} = 1 \quad \forall rc, \quad \sum_{m \in \Omega^F} x^{RC}_{m, dp^{RC}_{rc}, rc} = 1 \quad \forall rc \quad (1)$$

$$x^{RC}_{dp^{RC}_{rc}, m, rc} + \sum_{m' \in \Omega^F, m' \neq m} x^{RC}_{m', m, rc} = x^{RC}_{m, dp^{RC}_{rc}, rc} + \sum_{n \in \Omega^F, n \neq m} x^{RC}_{m, n, rc} \quad \forall rc, \forall m \quad (2)$$

$$\sum_{rc \in RC} \left( x^{RC}_{dp^{RC}_{rc}, n, rc} + \sum_{m \in \Omega^F, m \neq n} x^{RC}_{m, n, rc} \right) = 1 \quad \forall n \quad (3)$$

$$x^{RC}_{m, n, rc} + x^{RC}_{n, m, rc} \leq 1 \quad \forall rc, \forall m, \forall n \neq m \quad (4)$$

where (1) indicates an RC should start from its depot to perform the repairing task and return to its depot when works are over. (2) guarantees the sequential continuity of RC's repairing task. To recover the destroyed ADN to the pre-disaster state, (3) is stated to ensure that all faulted components must be visited (repaired) by RC. In addition, (3) also restricts that each faulted component can only be visited by one RC in the view of efficient repairing. (4) implies that there should be no cycles in the routing path of each RC.

The RC's arrival time at each faulted component is constrained accurately by constraints (5)-(7):

$$\begin{cases} AT^{RC}_{n, rc} \geq T^{MAX} - \left( x^{RC}_{dp^{RC}_{rc}, n, rc} + \sum_{m \in \Omega^F, m \neq n} x^{RC}_{m, n, rc} \right) M & \forall rc, \\ AT^{RC}_{n, rc} \leq T^{MAX} + \left( x^{RC}_{dp^{RC}_{rc}, n, rc} + \sum_{m \in \Omega^F, m \neq n} x^{RC}_{m, n, rc} \right) M & \forall n \end{cases} \quad (5)$$

$$\begin{cases} AT^{RC}_{n, rc} \geq T^{TR, RC}_{dp^{RC}_{rc}, n, rc} - \left( 1 - x^{RC}_{dp^{RC}_{rc}, n, rc} \right) M & \forall rc, \\ AT^{RC}_{n, rc} \leq T^{TR, RC}_{dp^{RC}_{rc}, n, rc} + \left( 1 - x^{RC}_{dp^{RC}_{rc}, n, rc} \right) M & \forall n \end{cases} \quad (6)$$

$$\begin{cases} AT^{RC}_{n, rc} \geq AT^{RC}_{m, rc} + T^{TR, RC}_{m, n, rc} + T^{RP}_{m, rc} - \left( 1 - x^{RC}_{m, n, rc} \right) M & \forall rc, \forall n \\ AT^{RC}_{n, rc} \leq AT^{RC}_{m, rc} + T^{TR, RC}_{m, n, rc} + T^{RP}_{m, rc} + \left( 1 - x^{RC}_{m, n, rc} \right) M & \forall m \neq n \end{cases} \quad (7)$$

where (5) indicates that if a faulted component is not assigned as a repair target for any RCs, it is considered that all RCs would not arrive at it until $T^{MAX}$ (not less than the time horizon of the entire restoration procedure). If the RC is assigned to repair faulted component *n* directly from its depot, as stated in (6), RC's arrival time at *n* would only be the travel time from its depot to *n*, whereas if this RC is assigned to repair another faulted component *m* prior to the task of repairing *n*, as formulated in (7), the RC's arrival time at *n* should be the sum of the repair ending time of *m* (i.e., the RC's arrival time at *m* plus the repair duration of *m*) and the travel duration between *m* and *n*.

At this point, the repair ending time of each faulted component can be calculated by the assigned RC's arrival time plus the repair duration:

$$t^{E, RP}_n = \min_{rc \in RC} \left[ AT^{RC}_{n, rc} + \left( x^{RC}_{dp^{RC}_{rc}, n, rc} + \sum_{m \in \Omega^F, m \neq n} x^{RC}_{m, n, rc} \right) T^{RP}_{n, rc} \right] \quad \forall n \quad (8)$$

The faulted component would be available for operation accordingly in the next time interval after the attained repair ending time. Hereupon, the status of each faulted component could be determined by disjunctive equation (9).

$$u^L_{n,t} = \begin{cases} 0 & \text{if } (t-1)\Delta T < t^{E, RP}_n \\ 1 & \text{if } (t-1)\Delta T \geq t^{E, RP}_n \end{cases} \quad \forall n, \forall t \quad (9)$$

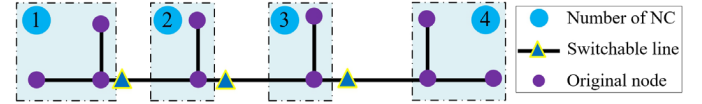

Fig. 5 Network reduction by aggregating nodes into cells.

***Network reduction technique***: By aggregating node blocks separated by sectionalizing switches in the original ADN into integrated node cells (NCs), a simplified graph $\mathcal{G}^{NC} = \{\mathcal{N}^{NC}, \mathcal{E}^{NC}\}$ is generated to facilitate the formulation of downscaled constraints. Where $\mathcal{N}^{NC}$ indicates the set of equivalent nodes representing each NC, and $\mathcal{E}^{NC}$ denotes the set of switchable lines among NCs. As an instance illustrated in Fig. 5, the 10-node physical network can be aggregated into 4 NCs according to the location of sectionalizing switches.

Toward this end, the fault clearance time of an NC depends on the final repair task in this cell, stated as:

$$u^{NC}_{nc, t} \leq \min_{n \in \Omega^{NC, F}_{nc}} u^L_{n, t} \quad \forall nc, \forall t \quad (10)$$

Note that by identifying the repair ending time of each faulted component precisely, the statuses of each faulted component and each NCs are formulated with VST in (9) and (10), respectively. Compared with the fixed time-step (FTS) model, VST-based formulations in RCDM are effective with any given optimization time interval $\Delta T$, and time checkpoints in (9) and (10) can be adjusted with the specific problem studied. More discussions on the applications of VTS and FTS in distribution system restoration can be found in [7].

*2) Operating crew dispatching model (OCDM)*

Similar to (1)-(4) in the RCDM, the routing sequence of each dispatched OC is formulated by constraints (11)-(14), where $oc \in OC$, $p, p', q \in \Omega^{MS}$.

$$\sum_{p \in \Omega^{MS}} x^{OC}_{dp^{OC}_{oc}, p, oc} = 1 \quad \forall oc, \quad \sum_{p \in \Omega^{MS}} x^{OC}_{p, dp^{OC}_{oc}, oc} = 1 \quad \forall oc \quad (11)$$

$$x^{OC}_{dp^{OC}_{oc}, p, oc} + \sum_{p' \in \Omega^{MS}} x^{OC}_{p', p, oc} = x^{OC}_{p, dp^{OC}_{oc}, oc} + \sum_{q \in \Omega^{MS}} x^{OC}_{p, q, oc} \quad \forall oc, \forall p \quad (12)$$

$$\sum_{oc \in OC} \left( x^{OC}_{dp^{OC}_{oc}, q, oc} + \sum_{p \in \Omega^{MS}} x^{OC}_{p, q, oc} \right) \leq 1 \quad \forall q \quad (13)$$

$$x^{OC}_{q, p, oc} + x^{OC}_{q, p, oc} \leq 1 \quad \forall oc, \forall p, \forall q \neq p \quad (14)$$

The OC's arrival time at each faulted component is constrained accurately by constraints (15)-(17) ($oc \in OC$, $p, p', q \in \Omega^{MS}$):

$$\begin{cases} AT_{q,oc}^{OC} \geq T^{MAX} - \left(x_{dp_{oc}^{OC},q,oc}^{OC} + \sum_{p \in \Omega^{MS}, p \neq q} x_{p,q,oc}^{OC}\right)M & \forall oc, \\ AT_{q,oc}^{OC} \leq T^{MAX} + \left(x_{dp_{oc}^{OC},q,oc}^{OC} + \sum_{p \in \Omega^{MS}, p \neq q} x_{p,q,oc}^{OC}\right)M & \forall q \end{cases} \quad (15)$$

$$\begin{cases} AT_{q,oc}^{OC} \geq T_{dp_{oc}^{OC},q,oc}^{TR,OC} - \left(1 - x_{dp_{oc}^{OC},q,oc}^{OC}\right)M & \forall oc, \\ AT_{q,oc}^{OC} \leq T_{dp_{oc}^{OC},q,oc}^{TR,OC} + \left(1 - x_{dp_{oc}^{OC},q,oc}^{OC}\right)M & \forall q \end{cases} \quad (16)$$

$$\begin{cases} AT_{q,oc}^{OC} \geq T_{p,oc}^{OC,S} + T_{p,oc}^{MSO} + T_{p,q,oc}^{TR,OC} - \left(1 - x_{p,q,oc}^{OC}\right)M & \forall oc, \forall q \\ AT_{q,oc}^{OC} \leq T_{p,oc}^{OC,S} + T_{p,oc}^{MSO} + T_{p,q,oc}^{TR,OC} + \left(1 - x_{p,q,oc}^{OC}\right)M & \forall p \neq q \end{cases} \quad (17)$$

Where (15) indicates that if an MS is not assigned as an operating task for any OCs, it is considered that all OCs would not arrive at it until $T^{MAX}$. If the OC is assigned to operating MS $q$ directly from its depot, as stated in (16), OC's arrival time at MS $q$ would only be the travel time from the depot to this MS, whereas if this OC is assigned to operate another MS $p$ prior to the task of operating MS $q$, as formulated in (16), OC's arrival time at MS $q$ should be the sum of the following three terms:

***a)*** The time when this OC arrives and can securely perform the operation of MS $p$;
***b)*** The duration required to operate the MS $p$;
***c)*** The travel duration between MS $p$ and MS $q$. As presented in equations (18)-(20), the above-mentioned time ***a*** should be the greater one between OC's arrival time at MS $p$ and the final fault clearing time for all NCs adjacent to MS $p$.
（Note that the latter term "the final fault clearing time for all NCs adjacent to MS $p$" is related to the repair ending time of each faulted component introduced in (8) in the RCDM, which reveals the *interdependency between the dispatching of OCs and RCs*. Hence, (20) is the coupling constraint between the RCDM and the OCDM.）

$$T_{p,oc}^{OC,S} = \max\left(AT_{p,oc}^{OC}, T_p^{NCR,MS}\right) \quad \forall oc, \forall p \quad (18)$$

$$T_p^{NCR,MS} = \max_{nc \in \Omega_p^{NCR,MS}} T_{nc}^{NCR} \quad \forall p \quad (19)$$

$$T_{nc}^{NCR} = \max_{n \in \Omega_{nc}^{NC,F}} t_n^{E,RP} \quad \forall nc \quad (20)$$

So far, the operating ending time of each MS can be calculated by the assigned OC's arrival time plus the operating duration:

$$t_q^{E,MSO} = \min_{oc \in OC}\left[T_{q,oc}^{OC,S} + \left(x_{dp_{oc}^{OC},q,oc}^{OC} + \sum_{p \in \Omega^{MS}, p \neq q} x_{p,q,oc}^{OC}\right)T_{q,oc}^{MSO}\right] \quad \forall q \quad (21)$$

The switchable line could be energized manually accordingly after the attained MS's operating ending time, shown in the VST-based disjunctive equation (22):

$$w_{l(q),t}^{MS} = \begin{cases} 0 & \text{if } (t-1)\Delta T < t_q^{E,MSO} \\ 1 & \text{if } (t-1)\Delta T \geq t_q^{E,MSO} \end{cases} \quad \forall q, \forall t \quad (22)$$

where $l(q)$ indicates the switchable line deployed with MS $q$.

Given the proposed RCDM and OCDM, the respective sequences of RC's and OC's tasks are guaranteed by the formulations of RC's and OC's arrival time formulated in (5)-(7) and (15)-(20). Different from the sequence of tasks, the priorities of tasks is not formulated by explicit expressions, but is attained by solving the global optimal solution of the proposed CPIRS optimization problem. That is, the significance of every task is jointly guaranteed by the proposed objective function and all the proposed constraints in this paper, and is not limited to the constraints formulated in the VRP. Whether one task should be prioritized over another depends on whether this scheduling strategy yields a smaller cumulative load-shedding cost, i.e., a more optimal objective function.

### B. Physical System Operation Model (PSOM)

*1) Constraints of load shedding*:

As there are no switchable lines inside an NC, the load demand inside this NC could not be safely supplied prior to the final fault of this NC being repaired:

$$\left(1 - u_{nc(i),t}^{NC}\right)P_{i,t}^{LD} \leq P_{i,t}^{Shed} \leq P_{i,t}^{LD} \quad \forall i \in N, \forall t \in T \quad (23)$$

where $nc(i)$ denotes the NC involving node $i$.

*2) Constraints of power output*:

Similarly, the power of substations and DERs (including GTs and RESs) inside an NC could only be safely generated after the final fault inside this NC is repaired. For $\forall t \in T$, the outputs of power sources should be constrained by:

$$\begin{cases} 0 \leq P_{i,t}^{S/GT} \leq u_{nc(i),t}^{NC} \overline{P}_{i,t}^{S/GT} \\ -u_{nc(i),t}^{NC} \overline{Q}_{i,t}^{S/GT} \leq Q_{i,t}^{S/GT} \leq u_{nc(i),t}^{NC} \overline{Q}_{i,t}^{S/GT} \end{cases} \quad \forall i \in N^{S/GT} \quad (24)$$

$$\begin{cases} 0 \leq P_{i,t}^{RES} \leq u_{nc(i),t}^{NC} \overline{P}_{i,t}^{RES} \\ -u_{nc(i),t}^{NC} \overline{Q}_{i,t}^{RES} \leq Q_{i,t}^{RES} \leq u_{nc(i),t}^{NC} \overline{Q}_{i,t}^{RES} \end{cases} \quad \forall i \in N^{RES} \quad (25)$$

The upward and downward ramping rate of the GTs is limited by the generator's capability. Furthermore, to ensure the safe operation of the system during the load restoration procedure, the spinning reserve capacity should be considered.

$$R_i^- \leq \left(P_{i,t}^{GT} - P_{i,t-1}^{GT}\right)/\rho_i^{GT} \leq R_i^+ \quad \forall i \in N^{GT}, \forall t \in T \setminus t_0 \quad (26)$$

*3) Constraints of power flow:*

The linearized DistFlow model, whose accuracy is validated by many cases of ADN's restoration [25], is used to impose the power flow constraints, stated as (for $\forall i \in N$ and $\forall t \in T$):

$$\begin{cases} \sum_{j \in U_i^N} P_{ij,t}^L - \sum_{j \in D_i^N} P_{ij,t}^L = P_{i,t}^{LD} - P_{i,t}^{Shed} - P_{i,t}^S - P_{i,t}^{GT} - P_{i,t}^{RES} \\ \sum_{j \in U_i^N} Q_{ij,t}^L - \sum_{j \in D_i^N} Q_{ij,t}^L = Q_{i,t}^{LD} - Q_{i,t}^{Shed} - Q_{i,t}^S - Q_{i,t}^{GT} - Q_{i,t}^{RES} \end{cases} \quad (27)$$

$$\begin{cases} -M\left(1 - w_{ij,t}\right) + \left(r_{ij}P_{ij,t}^L + x_{ij}Q_{ij,t}^L\right)/V_0 \leq V_{i,t} - V_{j,t} & \forall ij \in L \\ V_{i,t} - V_{j,t} \leq M\left(1 - w_{ij,t}\right) + \left(r_{ij}P_{ij,t}^L + x_{ij}Q_{ij,t}^L\right)/V_0 & \forall t \in T \end{cases} \quad (28)$$

The bus voltage and the line capacity are constrained by:

$$\underline{V}_i \leq V_{i,t} \leq \overline{V}_i \quad \forall i \in N, \forall t \in T \quad (29)$$

$$\begin{cases} -w_{l,t}\overline{P}_l^L \leq P_{l,t}^L \leq w_{l,t}\overline{P}_l^L \\ -w_{l,t}\overline{Q}_l^L \leq Q_{l,t}^L \leq w_{l,t}\overline{Q}_l^L \end{cases} \quad \forall l \in L, \forall t \in T \quad (30)$$

*4) Constraints of Frequency Response Rate (FRR):*

A simplified constraint [26] is adopted here to prevent potential unacceptable frequency deviations led by the disturbance bring by the load pickups. Concretely, newly restored loads in each time step should not exceed the maximum permitted restoration amount decided by the system frequency regulating ability, i.e., the value of the newly picked upload should be not greater than the total rated power of energized DERs multiplied by FRR.

$$\sum_{i\in N}\left(P_{i,t}^{LD}-P_{i,t}^{Shed}-P_{i,t-1}^{LD}+P_{i,t-1}^{Shed}\right)$$
$$\leq \sum_{i\in N}\left[u_{nc(i),t}^{NC}\left(\overline{P}_i^{GT}\gamma_i^{GT}+\overline{P}_i^{RES}\gamma_i^{RES}\right)\right] \quad \forall t \in T\setminus t_0 \quad (31)$$

*5) Constraints of radial operation:*

The ADN should be operated radially for the sake of secure operation. By abstracting the ADN into a simplified graph $\mathcal{G}^{NC}$ (described in Section II-A). Two conditions can guarantee the radiality of the network topology:

*R1)* Each subgraph (i.e., microgrid) is connected.

*R2)* The number of subgraphs (i.e., microgrids) equals the number of edges (i.e., switchable lines) plus the number of aggregated nodes (i.e., NCs).

To date, many ingenious methods are proposed to find connected graph separators (*R1*), such as the methods based on energization paths [25], the minimal spanning tree [38], and the single-commodity flow [39]. However, the tie lines (backup branches) are not considered in the energization path based method, and the minimal spanning tree methods could be ineffective (potential residual loops) when implemented in ADN with DERs [39]. Here the single-commodity flow is used to guarantee the ADN's connectivity *R1* (for $\forall t \in T$):

$$1-M\xi_{i',t} \leq \sum_{j'\in U_{i'}^{NC}} F_{i'j',t} - \sum_{j'\in D_{i'}^{NC}} F_{i'j',t} \leq 1+M\xi_{i',t} \quad \forall i' \in NC \quad (32)$$

$$-Mw_{i'j',t} \leq F_{i'j',t} \leq Mw_{i'j',t} \quad \forall i'j' \in \mathcal{E}^{NC} \quad (33)$$

where (32) indicates that single-commodity must flow out of each leaf node, and meanwhile, sufficient commodities should be generated at each root node to meet the commodity demand in each subgraph.

On this basis, *R2* will be satisfied accordingly by imposing the following constraint:

$$|NC| = \sum_{i'j'\in \mathcal{E}^{NC}} w_{i'j',t} + \sum_{i'\in \mathcal{N}^{NC}} \xi_{i',t} \quad (34)$$

*6) Constraints of the final network pattern:*

All microgrids (sub grids working in the islanding mode) are considered to be combined together at the end of the restoration to achieve steady energy supplies to customers. On the basis of the above formulations of radiality, the following constraint is stated to regulate the final network pattern:

$$\sum_{i'\in \mathcal{N}^{NC}} \xi_{i',t_{end}} = 1 \quad (35)$$

Where $t_{end}$ denotes the final time spot of the scheduling.

Constraint (35) enforces that only one root node is permitted in the final ADN, i.e., all microgrids are combined. If the root node is greater than 1, island(s) will still exist in the final network and the restoration process is not complete.

*7) Constraints of uncertain outputs of RESs:*

The predefined budget-constrained polyhedral uncertainty sets [40] are utilized to depict the uncertain outputs of RESs.

$$\overline{P}_{i,t}^{RES} = \hat{P}_{i,t}^{RES} + \left(\sigma_{i,t}^{+,RES}-\sigma_{i,t}^{-,RES}\right)\omega_i \hat{P}_{i,t}^{RES} \quad \forall i \in N^{RES}, \forall t \in T \quad (36)$$

$$\sum_{t\in T}\left(\sigma_{i,t}^{+,RES}+\sigma_{i,t}^{-,RES}\right) \leq B_i^{RES} \quad \forall i \in N^{RES} \quad (37)$$

$$\sigma_{i,t}^{-,RES}\in[0,1],\ \sigma_{i,t}^{+,RES}\in[0,1] \quad \forall t \in T \quad (38)$$

*C. Cyber System Availability Model (CSAM)*

As stated in (39), a cyber link is only available when every router involved in it is available, that is, only if every $u_{c',t}^C$ ( with the total number of $n_c^C$ ) takes 1, $u_{c,k,t}^{C,Link}$ can take 1. (40) means a router can be competent only when at least one cyber link is unblocked to communicate with the control centre.

$$u_{c,k,t}^{C,Link} \leq \sum_{c'\in L_{c,k}^C} u_{c',t}^C \Big/ n_{c,k}^{C,Link} \quad \begin{array}{l}\forall c \in \{C^{RCS},C^{GT},C^{Add}\},\\ \forall k \in [1,n_c^C], \forall t \in T\end{array} \quad (39)$$

$$u_{c,t}^C \leq M\sum_{k=1}^{n_c^C} u_{c,k,t}^{C,Link} \quad \forall c \in \{C^{RCS},C^{GT},C^{Add}\}, \forall t \in T \quad (40)$$

*D. Cyber-Physical Interdependency Model (CPIM)*

*1) Dependency of the cyber side on the physical side:*

The available energization, from the external power from the network or the local UPS, is a prerequisite for cyber routers to perform communication functions. Hereupon, for $\forall c \in \{C^{RCS},C^{Add},C^{GT},C^{RES},C^S\}$:

$$u_{c,t}^C \geq \max\left[\left(P_{i(c),t}^{LD}-P_{i(c),t}^{Shed}-\overline{P}_c^{Router}\right)\Big/M,\ u_{c,t}^{UPS,C}\right] \quad \forall t \in T \quad (41)$$

$$u_{c,t}^C \leq M\left(P_{i(c),t}^{LD}-P_{i(c),t}^{Shed}-\overline{P}_c^{Router}+u_{c,t}^{UPS,C}\right) \quad \forall t \in T \quad (42)$$

where $i(c)$ denotes the node corresponding to cyber router $c$.

The availability of the cyber router's power supply from the network depends on whether the actual power consumption of the router node is less than this cyber router's rated power consumption. The power supply from the local UPS is available when the time has not exceeded the rated power supply duration. For $\forall c \in \{C^{RCS},C^{Add},C^{GT},C^{RES},C^S\}$:

$$u_{c,t}^{UPS,C} = \begin{cases} 1 & \text{if } t\Delta T \leq T_c^{UPS,C} \\ 0 & \text{if } t\Delta T > T_c^{UPS,C} \end{cases} \quad \forall t \in T \quad (43)$$

*2) Dependency of the physical side on the cyber side:*

Each RCS can be operated remotely under two circumstances:

*a)* The cyber router controlling this RCS is available;

*b)* An external environment that supports secure operations, i.e., all faults are cleared in NCs connected by this RCS.

Considering the above two conditions and the remote operation time of the RCS, the remote closing status of each line can be formulated as (for $\forall s \in \Omega^{RCS}$, $\forall t \in T$):

$$w_{l(s),t}^{RCS} = \begin{cases} 1 & \text{if } u_{c(s),t}^C=1, \text{ and } (t-1)\Delta T \geq T_s^{NCR,RCS}+T_s^{RCSO} \\ 0 & \text{if } u_{c(s),t}^C=0, \text{ or } (t-1)\Delta T < T_s^{NCR,RCS}+T_s^{RCSO} \end{cases} \quad (44)$$

$$T_s^{NCR,RCS} = \max_{nc\in\Omega_s^{NC,RCS}} T_{nc}^{NCR} \quad \forall s \in \Omega^{RCS} \quad (45)$$

where $c(s)$ denotes the cyber router controlling the $s$-th RCS.

Note that MSs cannot be operated remotely:

$$w_{l(q),t}^{RCS} = 0 \quad \forall q \in \Omega^{MS}, \forall t \in T \quad (46)$$

where $l(q)$ indicates the switchable line deployed with MS $q$.

Considering both the remote-controlled mode and manual-controlled mode, the closing status of each sectionalizing switch can be eventually derived as:

$$\left(w_{l,t}^{MS}+w_{l,t}^{RCS}\right)\big/2 \leq w_{l,t} \leq w_{l,t}^{MS}+w_{l,t}^{RCS} \quad \forall l \in L, \forall t \in T \quad (47)$$

where $w_{l,t}$ will take 1 when one of $w_{l,t}^{MS}$ and $w_{l,t}^{RCS}$ takes 1.

*E. Two-Stage Robust Optimization Considering Uncertain Outputs of RESs:*

To ensure the priority power supply of critical loads, based on the resilience metric discussed in Section II-A, the load shedding costs based on expected energy not supplied (EENS) is chosen as the minimized objective function quantifying the reduction in system performance. Considering the uncertain outputs of RESs, this restoration scheduling problem is formulated as a two-stage robust optimization model. The objective function is stated as:

$$\min_{x}\left\{\max_{\sigma}\min_{y}\left(\sum_{i\in N}c_i^{Penalty}\sum_{t\in T}P_{i,t}^{Shed}\Delta T\right)\right\} \quad (48)$$

where $x$, $\sigma$, $y$ represent vectors of binary scheduling variables in the first stage, uncertain variables, and continuous operational variables, respectively. In the first stage, the system scheduler will act as a defender to make the optimal scheduling decisions $x$ under determined uncertainties $\sigma$. In the second stage, the uncertainties of RESs' outputs will play as an attacker to seek the worst restoration scenario $\sigma$ with a given optimal power flow $y$, adjusted by the system operation.

## IV. MODEL MODIFICATIONS AND SOLUTION METHODS

Several model modifications for boosting the solution and an effective solution algorithm will be introduced.

*A. Task Objective Clustering for RCs and OCs*

Currently, the tasks dispatched to each RC (or OC) are homogeneous and not customized, rendering potential combination explosions. The method proposed in [5] is adopted to cluster RCs' (OCs') tasks preliminarily (based on Euclidean distances between targets and depots) to narrow the feasible domain and boost the branching during the solution procedure.

*B. Model Linearization*

*1) Linearization for non-linear formulations:*

Since the uncertainties of RESs' outputs are taken into consideration, $\bar{P}^{RES}$ in (25) becomes a variable. Given intermediate variable $\alpha_{i,t}=u_{nc(i),t}^{NC}\bar{P}^{RES}$, the following two linearized equations are presented to replace (25):

$$0\leq P^{RES}\leq \alpha_{i,t} \quad \forall i\in N^{RES},\forall t\in T \quad (49)$$

$$\begin{cases}\alpha_{i,t}\geq\max\left\{0,\bar{P}^{RES}-M\left(1-u_{nc(i),t}^{NC}\right)\right\}\\ \alpha_{i,t}\leq\min\left\{\bar{P}^{RES},Mu_{nc(i),t}^{NC}\right\}\end{cases}\forall i\in N^{RES},\forall t\in T \quad (50)$$

*2) Linearization for disjunctive expressions:*

*a)* Given the intermediate binary variable $\beta_{n,rc}^{RC}$ satisfying $\sum_{rc\in RC}\beta_{n,rc}^{RC}=1$ ($\forall n\in\Omega^F$) and $\beta_{q,oc}^{OC}$ satisfying $\sum_{rc\in RC}\beta_{q,oc}^{OC}=1$ ($\forall q\in\Omega^{MS}$), the following two linearized equations are raised to replace the minimal selections (8) and (21):

$$\begin{cases}\tau_{n,rc}^{RC}-T^{MAX}\left(1-\beta_{n,rc}^{RC}\right)\leq t_n^{E,RP}\leq\tau_{n,rc}^{RC}\\ \tau_{n,rc}^{RC}=AT_{n,rc}^{RC}+\left(x_{dp_{rc}^{RC},n,rc}^{RC}+\sum_{m\in\Omega^F,m\neq n}x_{m,n,rc}^{RC}\right)T_{n,rc}^{RP}\end{cases}\forall n,\forall rc \quad (51)$$

$$\begin{cases}\tau_{q,oc}^{OC}-T^{MAX}\left(1-\beta_{q,oc}^{OC}\right)\leq t_q^{E,MSO}\leq\tau_{q,oc}^{OC}\\ \tau_{q,oc}^{OC}=T_{q,oc}^{OC,S}+\left(x_{dp_{oc}^{OC},q,oc}^{OC}+\sum_{p\in\Omega^{MS},p\neq q}x_{p,q,oc}^{OC}\right)T_{q,oc}^{MSO}\end{cases}\forall q,\forall oc \quad (52)$$

*b)* Let $\varepsilon$ be a sufficiently small positive number and $M$ be a sufficient big positive number, then disjunctive expressions (9), (22) and (43) can be respectively linearized as:

$$(t-1)\Delta T-t_n^{E,RP}\leq u_{n,t}^L M-\varepsilon\leq M+(t-1)\Delta T-t_n^{E,RP}$$
$$\forall\varepsilon<\left|\Delta T-t_n^{E,RP}\right|,\forall n\in\Omega^F,\forall t\in T \quad (53)$$

$$(t-1)\Delta T-t_q^{E,MSO}\leq w_{q,t}^{MS}M-\varepsilon\leq M+(t-1)\Delta T-t_q^{E,MSO}$$
$$\forall\varepsilon<\left|\Delta T-t_q^{E,MSO}\right|,\forall q\in\Omega^{MS},\forall t\in T \quad (54)$$

$$\left(T_c^{UPS,C}-t\Delta T+\varepsilon\right)/M\leq u_{c,t}^{UPS,C}\leq 1+\left(T_c^{UPS,C}-t\Delta T+\varepsilon\right)/M$$
$$\forall\varepsilon<\left|T_c^{UPS,C}-\Delta T\right|,\forall q\in\Omega^{MS},\forall t\in T \quad (55)$$

where $\varepsilon$ is introduced to deal with equalities of $t\Delta T=t_n^{E,RP}$, $t\Delta T=t_q^{E,MSO}$ and $T_c^{UPS,C}=t\Delta T$.

*c)* For $\forall s\in\Omega^{RCS}$, $\forall t\in T$, the disjunctive expression (44) can be linearized by (56). Note that although (56) and (44) are not exactly equivalent, this alternative is reliable for the studied problem because (56) gives the accurate upper bound of $w_{l(s),t}^{RCS}$.

$$w_{l(s),t}^{RCS}\leq\min\left\{u_{c(s),t}^C,\ 1+\left[(t-1)\Delta T-T_s^{NCR,SS}-T_s^{RCSO}\right]/M\right\} \quad (56)$$

At this point, the proposed mathematical optimization model has been reformulated in a monolithic form of MILP.

*C. Model Adjustment for Expanding OCs' Targets*

To tackle this thorny issue and make the scheduling of OC more flexible, task objects of OCs in the proposed OCDM are designated as both RCSs and MSs. Since the remote operation of RCSs is always more efficient than the manual operation, this modification would not affect the optimality of the solution.

*D. Solution Algorithm*

The proposed two-stage robust cyber-physical restoration scheduling model can be rewritten with compact matrices:

$$\min_{x}\left\{\max_{\sigma}\min_{y}\boldsymbol{b}^{\mathrm{T}}\boldsymbol{y}\right\} \quad (57)$$

s.t. [I] $\boldsymbol{Ax}\leq\boldsymbol{d}$, [II] $\boldsymbol{Dx}+\boldsymbol{Fy}\leq\boldsymbol{f}$, [III] $\boldsymbol{G\sigma}+\boldsymbol{Hy}\leq\boldsymbol{g}$ (58)

where [I] summaries (1)-(7), (10)-(20), (40), (39), (52) and (54). [II] contains (23)-(24), (26)-(34), (42), (45)-(47), (55) and (56). [III] represents (36)-(38).

The column-and-constraint generation (C&CG) algorithm is customized to solve the proposed robust restoration scheduling model. The solving procedure is presented below.

**Algorithm 1: customized C&CG algorithm.**
**Step 1.** Set $LB=0$, $UB=+\infty$, $r=0$, convergence tolerance $\varepsilon$.
**Step 2.** Solve the master problem (MP), derive the optimal scheduling scheme $\boldsymbol{x}^{r*}$ and the optimal objective $\mu^{r*}$, and update the lower bound $LB=\max\left\{\mu^{r*},LB\right\}$.

$$(\mathbf{MP})\ \min_{x}\mu \quad (59)$$

$$\text{s.t.}\ \boldsymbol{Ax}\leq\boldsymbol{d},\ \mu\geq 0 \quad (60)$$

**Step 3.** Fix $\boldsymbol{x}^{r*}$, obtain the following subproblem (SP) by taking the dual of the second stage program according to the strong duality, solve this SP, and derive the optimal objective $\vartheta(x^{r*})^*$ of this SP.

$$\text{(SP)} \max_{\sigma} \vartheta(x^{r*}) = (f - Dx^{r*})^T \pi_1 + (g - G\sigma)^T \pi_2 \quad (61)$$

$$s.t., \ F^T \pi_1 + H^T \pi_2 = b, \ \pi_1 \leq 0, \ \pi_2 \leq 0 \quad (62)$$

1) If $\vartheta(x^{r*})^* < +\infty$, obtain the worst scenario $\sigma^{r*}$ and the optimal operational scheme $y^{r*}$, update the upper bound $UB = \min\{\vartheta(x^{r*}), UB\}$. If $1 - LB/UB < \varepsilon$, return $x^* = x^{r*}$ and **stop**. Otherwise, update $r = r+1$, create new variables $y^r$, go to **Step 2** and add the following constraints to the MP.

$$\mu \geq b^T y^r, \ Dx + Fy^r \leq f, \ G\sigma^{r*} + Hy^r \leq g \quad (63)$$

2) If $\vartheta(x^{r*})^* = +\infty$, derive the identified scenario $\sigma^{r*}$. If $1 - LB/UB < \varepsilon$, return $x^* = x^{r*}$ and **stop**. Otherwise, update $r = r+1$, create new $y^r$, go to **Step 2** and add the following constraints to the MP.

$$Dx + Fy^r \leq f, \ G\sigma^{r*} + Hy^r \leq g \quad (64)$$

**Note:** Linearization of $\sigma^T \pi_2$ in (61) and $\sigma^T \pi_2 = \sum_{s'} \sigma_{s'} \pi_{2,s'}$. First, fit each continuous variable $\sigma_{s'}$ with a polynomial:

$$\begin{cases} \sigma_{s'} = \underline{\sigma}_{s'} + \eta_{s'}(2^0 \varsigma_{s',0} + 2^1 \varsigma_{s',1} + \cdots + 2^{n_{s'}} \varsigma_{s',n_{s'}}) \\ 0 \leq \eta_{s'}(2^0 \varsigma_{s',0} + 2^1 \varsigma_{s',1} + \cdots + 2^{n_{s'}} \varsigma_{s',n_{s'}}) \leq \overline{\sigma}_{s'} - \underline{\sigma}_{s'} \end{cases} \quad (65)$$

where $\varsigma_{s',0}, \varsigma_{s',1}, \ldots, \varsigma_{s',n_{s'}}$ denotes binary variables, and $\eta_{s'}$ represents the fit precision factor. Each bilinear item $\sigma_{s'} \pi_{2,s'}$ can be linearized as:

$$\begin{cases} \sigma_{s'} \pi_{2,s'} = \underline{\sigma}_{s'} \pi_{2,s'} + \eta_{s'}(2^0 \varsigma_{s',0} + 2^1 \varsigma_{s',1} + \cdots + 2^{n_{s'}} \varsigma_{s',n_{s'}}) \\ -M \varsigma_{s',k'} \leq \upsilon_{s',k'} \leq M \varsigma_{s',k'} \\ \pi_{2,s'} - M(1 - \varsigma_{s',k'}) \leq \upsilon_{s',k'} \leq \pi_{2,s'} + M(1 - \varsigma_{s',k'}) \end{cases} \quad (66)$$

where $\upsilon_{s',k'}$ is the intermediate variable to denote $\varsigma_{s',k'} \pi_{2,s'}$.

## V. CASE STUDIES

The IEEE 123-node system [41], shown in Fig. 6, is used to verify the effectiveness of the proposed CPIRS method. This test feeder is operated with a nominal voltage of 4.16 kV and supplies the total original load demand of 3.49 MW. Overall, 11 sectionalizing switches, including 5 RCSs and 6 MSs, are placed on switchable lines, and therefore the ADN is divided into 10 NCs. It is assumed that the load in each NC is nearly three-phase balanced to mitigate the network losses, which can be addressed by the previous network planning [13]. The remote operating duration of all switches is set as 2 min. It is considered that 8 lines are damaged after an extreme event. Expected manual operating durations of switches and expected repair durations of faulted lines are referred to [25]. Overall, 3 GTs (each with a capacity of 200 kW) and 3 RESs (each with a capacity of 300 kW) are considered as DERs. The FRR of each DER is set as 5% [26]. For RES uncertainties, the maximum prediction error, denoted as $\omega_i$ in (36), is set as 30% [9], and the budget of uncertainty, denoted as $B_i^{RES}$ in (37), is set as 5.

The time interval and the time horizon of the restoration schedule are set as 30 and 450 minutes, respectively. Predicted fluctuations of maximum predicted load demand and RESs of the 15 time slots are referred to [9]. Overall, 1.27 MW loads, located at 22 nodes and 7 NCs, are considered as critical loads. The penalty cost of the common load and the critical load is set as $14/kWh and $1000/kWh, respectively. Two RCs and one OC are deployed at three depots. The travel time is calculated is obtained by dividing the distance between task points (assuming as twice the straight-line distance) on the map by the preset traveling speed of vehicles. To be conservative, the travel speed of all crews is set as 5 km/h considering traffic jams.

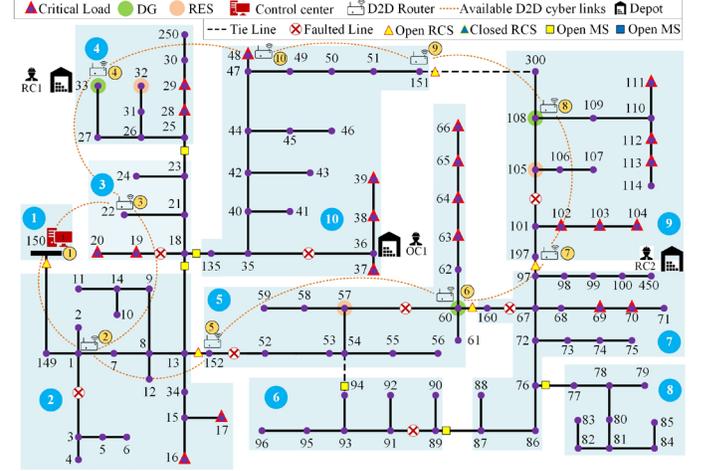

Fig. 6 The IEEE 123 node systems with restoration resources.

A D2D cyber network is established with 10 cyber routers, including 7 routers deployed at each FTUs controlling RCS and each node cell with DERs, and 3 added relay routers. "Cisco 1240 Connected Grid Router" [42] is taken as the real-life wireless cyber router in this paper. The power consumption of a router is set to 75W. The coverage distance of the cyber router is 1,200 meters, and to be on the safe side, 1,000 meters is set as the maximum secure communication distance for each cyber router. Considering the post-disaster communication pressure, the UPS time with the maximum power consumption supported by the integrated modular battery backup unit (BBU) is about 300 minutes. This conservative UPS time is considered in case studies to test the effectiveness of the proposed cyber-physical interdependent restoration scheduling (CPIRS) model. Note that in practical applications, the specific UPS time can be set more accurately by the experience of the equipment supplier.

The convergence tolerance of the C&CG algorithm is set as 0.1%. All simulations are performed by Gurobi 9.5.1 [45] on a PC with a Core i7-7700 CPU and 16GB RAM.

### A. Analysis of Optimal Restoration Scheduling Scheme

Prior to the solution, by implementing the task clustering method, two RCs are assigned 4 repair tasks each based on the distance between targets and their depots. Whereafter, the proposed CPIRS model is solved in 492s. The obtained optimal scheduling schemes of crew dispatching are presented in Table I, and the final network topology and scheduling paths of crews are intuitively illustrated in Fig. 7. It can be seen that two RCs and one OC are well coordinated to conduct the restoration task. Since the goal of the proposed model is to reduce the cumulative EENS over the whole system, neither RC is assigned preferentially to repair the closest faulted line, even if such travel time is the most economical. Note that on account of the interdependencies between OCDM and RCDM, although OC1 reached the faulty line at $t = 11$ min, it has to wait there for a secure operational condition until $t = 242$ min.

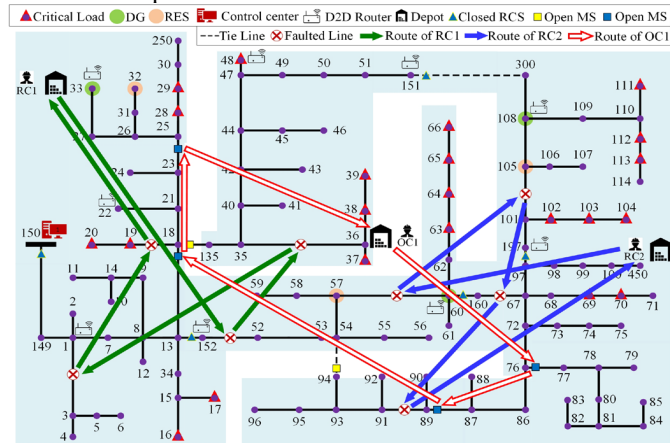

Fig. 7 The Final network topology and scheduling paths for RCs and OC.

TABLE I SCHEDULING TIME POINTS AND PATHS OF RCS AND OC

| Crews | Arrival/leaving time (min)@[Visiting location] |
|---|---|
| RC1 | [Depot-RC1] → **10/63**@[54,57] → **70/182**@[35,36] → **192/290**@[1,3] → **295/386**@[18,19] → [Depot-RC1] |
| RC2 | [Depot-RC2] → **7/50**@[57,60] → **60/109**@[101,105] → **112/232**@[67,160] → **240/334**@[89,91] → [Depot-RC2] |
| OC1 | [Depot-OC1] → **11/242**@[76,77] → **247/343**@[87,89] → **358/394**@[13,18] → **403/413**@[23,25] → [Depot-OC1] |

TABLE II MILESTONES IN THE CYBER-PHYSICAL ADN'S RESTORATION

| Time (min) | Significant energizing (**E**) & de-energizing (**D**) events | | Closing Switches | |
|---|---|---|---|---|
| | Cyber router- x | NC- x | Remotely [Activated router] | Manually |
| 0 | (**E**) 1, 4 by MG, others by UPS | (**E**) 1,4 | | |
| 63 | (**E**) 5, 6 by MG | (**E**) 5 | | |
| 109 | (**E**) 7, 8 by MG | (**E**) 9 | | |
| 184 | (**E**) 11, 12 by MG | (**E**) 10 | (151,300) [**151**] | |
| 234 | | (**E**) 7 | (60,160) [**60**] (97,197) [**197**] | |
| 242 | | (**E**) 8 | | (76,77) |
| 292 | (**E**) 2 by MG | (**E**) 2 | (149,150) [**150**] (13,152) [**152**] | |
| 300 | (**D**) 3 (Relay router) | | | |
| 343 | | (**E**) 6 | | (87,89) |
| 394 | (**E**) 3 by MG | (**E**) 3 | | (13,18) |
| 413 | | | | (23,25) |

Milestones in this CPIRS are summarized in Table II. The cyber-physical interdependency can be found running throughout the entire restoration scheduling procedure. Except for NC-1 and NC-4, the network falls into a blackout at first, and all cyber routers are powered up by their UPS. Microgrids (MGs) are generated as RCs clear the faults in NC-5 and NC-9, allowing loads and cyber routers to be energized. At $t = 184$ min, the backup tie line (151, 300) timely contributes to the power transfer for NC-10. Similarly, at $t = 234$ min, two RCSs are closed remotely through the available D2D cyber links between the control centre and energized cyber routers, and NC-7 is energized thereupon. Next, two RCSs and one MS are closed to energize NC-7 and NC-8. At $t = 292$ min, just before cyber routers' UPS runs out of power at $t = 300$ min, the remaining two RCSs are closed remotely to energize NC-2 via available D2D cyber links. Finally, three MSs are operated by OC to supply power to NC-6 and NC-3. So far, all NCs are energized and all microgrids are connected together to achieve a thorough recovery at the end of CPIRS.

### B. Benchmarking of the Proposed Method

Two groups of comparative cases focusing on the physical side and the cyber side will be studied respectively to verify the superiorities of the proposed CPIRS method.

*1) Microgrids: Tapping the potential of DERs*

The impact of DERs to construct microgrids will be tested. Methods proposed most recently in [13] and [14], in which the cyber-physical interdependent restoration scheme is emphasized but DERs are not considered, are set as Benchmark-I. The proportion of restored critical loads in each scheduling period (i.e., the ratio of the restored critical load to the total load in the corresponding period), and the accumulated penalty cost of load shedding in each time period are presented in Fig. 8 and Fig. 9, respectively.

In the cyber-physical interdependent ADN, timely established microgrids based on DERs can also contribute greatly in the early stage of restoration, which is intuitively reflected in the rapid energization of critical loads in the first ten scheduling periods. Compared with Benchmark-I, the proposed CPIRS method can ultimately save up to 32.8% ($1.65 million) load shedding costs. Note that the phenomenon that the proportion of restored critical load may reduce slightly along with the time passed is due to the fluctuations of RESs' outputs are considered in this paper. Certain critical loads whose power supplies totally depend on wind power may be shed in some time period.

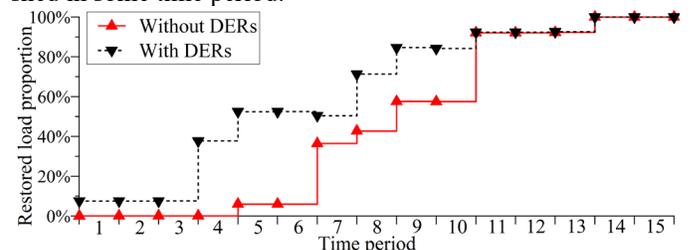

Fig. 8 Proportion of restored critical load in each scheduling time period with or without DERs.

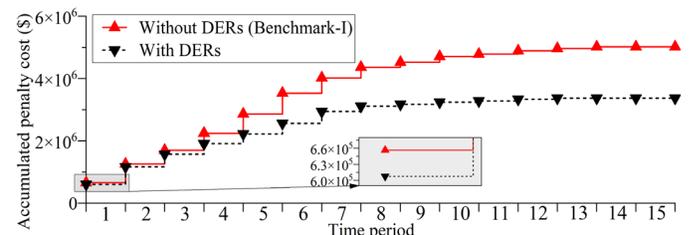

Fig. 9 Accumulated penalty costs of load shedding in each scheduling time period with or without DERs.

*2) Communications: With and Without D2D Communication*

In most prevalent publications [5]-[9], [25], [26] regarding ADN restoration, DERs are well considered as potential power supplies, but the cyber-physical interdependencies are ignored, (i.e., cyber systems are assumed to survive a disaster perfectly). In this case, some tasks of the original decision scheme may become infeasible due to a lack of communication. The methods provided in [5]-[9], [25], [26] are hence set as Benchmark-II. We assume that in Benchmark-II, the cyber infrastructure will be completely recovered at 300 min (i.e., without D2D communication). Prior to this time, tasks that do not require remote controls can be performed. Note that it is assumed that the control centre is aware of the routers and the time periods losing communications. If the dispatch centre is blind to when and where communications are lost, some crews may be stuck in long waits according to the original scheduling schemes, and thereby the results will only be worse.

Oppositely, post-disaster ad hoc cyber networks can be set up instantly by the proposed CPIRS via ad hoc wireless communication (D2D communication), so the availability of communication can be obtained in the early stages after a disaster, and maintenance resources can be well-coordinated with different UPS time of each D2D router. Two cases with UPS time of 180 minutes and 300 minutes are set to compare with Benchmark-II. Corresponding results are presented in Fig. 10 and Fig. 11.

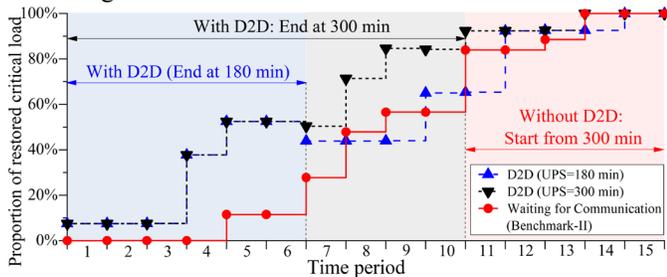

Fig. 10 Proportion of restored critical load in each scheduling time period with various communication availabilities.

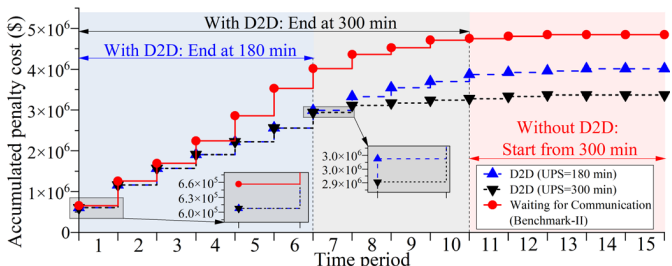

Fig. 11 Accumulated penalty costs of load shedding in each scheduling time period with various communication availabilities.

As per these two figures, before the end of the 6th time period (i.e., before 180 minutes), both two cases with pre-availabilities of D2D communication behave the same. Whereas in the next four time periods, the case with UPS=300 min outperforms the one with UPS=180 min on rapid critical load restoration. The restored critical load ratio of Benchmark-II did not reach a similar level to that of two cases with pre-availabilities of D2D communication until the 11th time period (after the communication is regained). Both two cases with D2D communication outperform Benchmark-II because the remote operation of switches can only be conducted through the established cyber network, which is significant for the loads that cannot obtain the power from DERs and have to access the upper-level grid. It can be seen that the establishment of communications in the early stages of a disaster is of vital significance to service restoration. Specifically, taking the case with a UPS time of 300 minutes as an example, compared with Benchmark-II, the proposed CPIRS method can ultimately save up to 30.5% ($1.48 million) load shedding costs.

*C. Discussion of RES Uncertainties*

The budget-constrained polyhedral uncertainty set, stated in equations (32)-(34), consists of three parts to characterize the uncertainty: 1) The forecasting power outputs of the wind turbine. 2) The MPE of each time period. 3) The BoU equaling the sum of prediction errors over the whole restoration horizon.

Even considering the post-disaster scenarios, the accuracy of the first part, i.e., the forecasting power outputs of the wind turbine, is being continuously improved by state-of-the-art prediction methods, such as the data-driven based model [43] and the model-data hybrid-driven model [44]. In addition, the restoration scheduling can only begin right after the disaster, instead of during the disaster. Therefore, the difficulty of the wind power forecasting focused by the restoration process could be relatively lower than that of the pre-disaster prediction.

However, any prediction method cannot be completely accurate. For the utilized budget-constrained polyhedral uncertainty set, the MPE and the BoU, which characterize the uncertainty level of wind power outputs, can be customized for the studied ADN. Here, the sensitivity analyses of these two parameters are implemented to demonstrate the effectiveness of the utilized budget-constrained polyhedral uncertainty set.

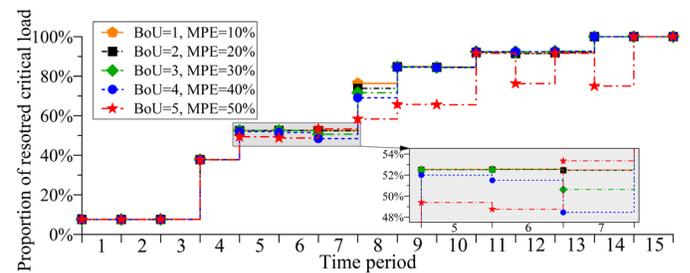

Fig. 12 Proportion of restored critical load in each scheduling period under various RES uncertainty levels.

It can be seen from Fig. 12 that the proportion of restored critical load in some scheduling time period reduces as the uncertainty of RESs' outputs climbs up. To attain an executable scheduling scheme, the proposed robust CPIRS model will find the worst scenario with uncertainties. Wind power may face a desperate shortage in some time periods when the uncertainty level climbs up severely, resulting in the lack of power supply of critical loads. Numerical results remind the significance of considering the uncertainty pattern of RESs' outputs.

Generally speaking, by setting the MPE and the BoU conservatively, the wind power outputs in the proposed robust restoration scheduling model can be formulated by the budget-constrained polyhedral uncertainty sets.

## VI. CONCLUSIONS

This paper proposes an ad hoc wireless communication assisted CPIRS framework for ADN restoration, where repair crews, operating crews, and system operations with DERs are all well-coordinated under cyber-physical interdependencies. Considering uncertain outputs of RESs, an MILP-based two-stage robust optimization model is established after the linearization. The C&CG algorithm is customized to solve this proposed model. Numerical results show that by implementing our method, microgrids can be formed efficiently by the instantly established wireless D2D cyber network to restore critical loads in time. Overall, 32.8% and 30.5% load shedding costs can be reduced compared to prevalent methods that ignore DERs and cyber-physical interdependencies, respectively.

Three issues deserve further research in the future: 1) The detailed modelling of the transportation system coupling with the restoration scheduling. 2) The detailed modelling of physical/mathematical properties of the wireless communication system. 3) More accurate models to represent the uncertainty pattern of RESs' outputs right after the disaster.

REFERENCES

[1] C. Chen, J. Wang and D. Ton, "Modernizing distribution system restoration to achieve grid resiliency against extreme weather events: an integrated solution," *Proc. IEEE*, vol. 105, no. 7, pp. 1267-1288, Jul. 2017.
[2] M. Panteli, D. N. Trakas, P. Mancarella and N. D. Hatziargyriou, "Power systems resilience assessment: hardening and smart operational enhancement strategies," *Proc. IEEE*, vol. 105, no. 7, pp. 1202-1213, Jul. 2017.
[3] M. Eriksson, M. Armendariz, O. O. Vasilenko, A. Saleem and L. Nordström, "Multiagent-based distribution automation solution for self-healing grids," *IEEE Trans. Ind. Electron.*, vol. 62, no. 4, pp. 2620-2628, Apr. 2015.
[4] C. Wang, K. Pang, M. Shahidehpour, F. Wen, H. Ren and Z. Liu, "Flexible joint planning of sectionalizing switches and tie lines among distribution feeders," *IEEE Trans. Power Syst.*, vol. 37, no. 2, pp. 1577-1590, Mar. 2022.
[5] A. Arif, Z. Wang, J. Wang and C. Chen, "Power distribution system outage management with co-optimization of repairs, reconfiguration, and DG dispatch," *IEEE Trans. Smart Grid*, vol. 9, no. 5, pp. 4109-4118, Sep. 2018.
[6] S. Lei, C. Chen, Y. Li and Y. Hou, "Resilient disaster recovery logistics of distribution systems: co-optimize service restoration with repair crew and mobile power source dispatch," *IEEE Trans. Smart Grid*, vol. 10, no. 6, pp. 6187-6202, Nov. 2019.
[7] B. Chen, Z. Ye, C. Chen, and J. Wang, "Toward a MILP modeling framework for distribution system restoration," *IEEE Trans. Power Syst.*, vol. 34, no. 3, pp. 1749–1760, May 2019
[8] T. Ding, Z. Wang, W. Jia, B. Chen, C. Chen, and M. Shahidehpour, "Multiperiod distribution system restoration with routing repair crews, mobile electric vehicles, and soft-open-point networked microgrids," *IEEE Trans. Smart Grid*, vol. 11, no. 6, pp. 4795–4808, Nov. 2020.
[9] H. Wu, Y. Xie, Y. Xu, Q. Wu, C. Yu, and J. Sun, "Robust coordination of repair and dispatch resources for post-disaster service restoration of the distribution system," *Int. J. Electr. Power Energy Syst.*, vol. 136, p. 107611, May 2021.
[10] S. Paul, F. Ding, K. Utkarsh, W. Liu, M. J. O'Malley and J. Barnett, "On vulnerability and resilience of cyber-physical power systems: a review," *IEEE Syst. J.*, Early access, doi: 10.1109/JSYST.2021.3123904.
[11] Report on the 2017 Atlantic Hurricane Season's impact on communications. [Online]. Available: https://docs.fcc.gov/public/attachments/DOC-353805A1.pdf
[12] B. Ti, G. Li, M. Zhou and J. Wang, "Resilience assessment and improvement for cyber-physical power systems under typhoon disasters," *IEEE Trans. Smart Grid*, vol. 13, no. 1, pp. 783-794, Jan. 2022.
[13] X. Liu, B. Zhang, B. Chen, A. Aved, and D. Jin, "Towards optimal and executable distribution grid restoration planning with a fine-grained power-communication interdependency model," *IEEE Trans. Smart Grid*, Early access, Feb. 2022, doi: 10.1109/tsg.2022.3149973.
[14] Z. Ye, C. Chen, R. Liu, K. Wu, Z. Bie, G. Lou, W. Gu and Y. Yuan, "Boost distribution system restoration with emergency communication vehicles considering cyber-physical interdependence," *IEEE Trans. Smart Grid*, 2022. Early access, doi: 10.1109/TSG.2022.3205039.
[15] Z. Li, M. Shahidehpour, F. Aminifar, A. Alabdulwahab and Y. Al-Turki, "Networked microgrids for enhancing the power system resilience," *Proc. IEEE*, vol. 105, no. 7, pp. 1289-1310, Jul. 2017.
[16] Y. Kabalci, "A survey on smart metering and smart grid communication", *Renew. Sustain. Energy Rev.*, vol. 57, pp. 302-318, May 2016.
[17] M. Erol-Kantarci and H. T. Mouftah, "Energy-efficient information and communication infrastructures in the smart grid: A survey on interactions and open issues," *IEEE Commun. Surveys Tuts.*, vol. 17, no. 1, pp. 179-197, Mar. 2015.
[18] S. Pincetl, "Cities are not isolates: To reduce their impacts a change in urban-rural interdependencies and the direction of modernity are required," *Advances in Applied Energy*, vol. 7, p. 100104, Sept. 2022.
[19] P. P. Parikh, M. G. Kanabar and T. S. Sidhu, "Opportunities and challenges of wireless communication technologies for smart grid applications," *IEEE PES General Meeting*, Jul. 2010, pp. 1-7.
[20] S. Alam, M. F. Sohail, S. A. Ghauri, I. M. Qureshi, and N. Aqdas, "Cognitive radio based smart grid communication network," *Renew. Sustain. Energy Rev.*, vol. 72, pp. 535-548, May 2017.
[21] A. Aijaz, "Private 5G: the future of industrial wireless," *IEEE Ind. Electron. Mag.*, vol. 14, no. 4, pp. 136-145, Dec. 2020.
[22] W. Miao, L. Wei, C. Jiang, B. Guo, W. Li, L. Wang, R. Liu and J. Zou, "Coverage and capacity analysis of LTE-based power wireless private network," *2017 11th IEEE International Conference on Anti-counterfeiting, Security, and Identification (ASID)*, Oct. 2017, pp. 119-124.
[23] W. Ji, M. Zhang, Y. Wang, Z. Huang and J. Weng, "Research and application of smart distributed feeder automation based on 5G wireless communication," *2021 China International Conference on Electricity Distribution (CICED)*, Apr. 2021, pp. 460-463.
[24] K. Ali, H. X. Nguyen, Q. -T. Vien, P. Shah and Z. Chu, "Disaster management using D2D communication with power transfer and clustering techniques," *IEEE Access*, vol. 6, pp. 14643-14654, Jan. 2018.
[25] G. Zhang, F. Zhang, X. Zhang, K. Meng, and Z. Y. Dong, "Sequential disaster recovery model for distribution systems with co-optimization of maintenance and restoration crew dispatch," *IEEE Trans. Smart Grid*, vol. 11, no. 6, pp. 4700-4713, Nov. 2020.
[26] B. Chen, C. Chen, J. Wang, and K. L. Butler-Purry, "Multi-time step service restoration for advanced distribution systems and microgrids," *IEEE Trans. Smart Grid*, vol. 9, no. 6, pp. 6793–6805, Nov. 2018.
[27] Q. Wang, G. Zhang, and F. Wen, "A survey on policies, modelling and security of cyber‐physical systems in smart grids," *Energy Convers. Econ.*, vol. 2, no. 4, pp. 197-211, Dec. 2021.
[28] C. Wang, K. Pang, M. Shahidehpour and F. Wen, "MILP-based fault diagnosis model in active power distribution networks," *IEEE Trans. Smart Grid*, vol. 12, no. 5, pp. 3847-3857, Sept. 2021.
[29] G. Zhang, X. Tong, Q. Hong, X. Lu and C. D. Booth, "A novel fault isolation scheme in power system with dynamic topology using wide-area information," *IEEE Trans. Ind. Inform.*, vol. 18, no. 4, pp. 2399-2410, Apr. 2022.
[30] K. Miranda, A. Molinaro and T. Razafindralambo, "A survey on rapidly deployable solutions for post-disaster networks," *IEEE Commun. Mag.*, vol. 54, no. 4, pp. 117-123, April 2016.
[31] M. Wang and Z. Yan, "A survey on security in D2D communications," *Mob. Networks Appl.*, vol. 22, no. 2, pp. 195–208, Apr. 2017.
[32] P. Ge, F. Teng, C. Konstantinou, and S. Hu, "A resilience-oriented centralised-to-decentralised framework for networked microgrids management," *Appl. Energy*, vol. 308, p. 118234, Feb. 2022.
[33] J. Z. Moghaddam, M. Usman and F. Granelli, "A Device-to-Device communication-based disaster response network," *IEEE Trans. Cogn. Commun. Netw.*, vol. 4, no. 2, pp. 288-298, Jun. 2018.
[34] W. Liu, Q. Gong, H. Han, Z. Wang and L. Wang, "Reliability modeling and evaluation of active cyber physical distribution system," *IEEE Trans. Power Syst.*, vol. 33, no. 6, pp. 7096-7108, Nov. 2018.
[35] B. Bhattarai, L. Marinovici, F. Tuffner, K. Schneider, X. Fan, F. Rutz and G. Kandaperumal, "Prototypical communication systems for electrical distribution system analysis: Design basis and exemplification through co-simulation, " *IET Smart Grid*, vol. 5, no. 5, pp. 363-379, Jun. 2022.
[36] Z. Chen and M. Kountouris, "Decentralized opportunistic access for D2D underlaid cellular networks," *IEEE Trans. Commun.*, vol. 66, no. 10, pp. 4842-4853, Oct. 2018.
[37] K. Braekers, K. Ramaekers, and I. Van Nieuwenhuyse, "The vehicle routing problem: state of the art classification and review, " *Comput. Ind. Eng.*, vol. 99, pp. 300-313, Sep. 2016.
[38] M. E. Baran and F. F. Wu, "Network reconfiguration in distribution systems for loss reduction and load balancing," *IEEE Trans. Power Del.*, vol. 4, no. 2, pp. 1401-1407, April 1989.
[39] Y. Wang, Y. Xu, J. Li, J. He and X. Wang, "On the radiality constraints for distribution system restoration and reconfiguration problems," *IEEE Trans. Power Syst.*, vol. 35, no. 4, pp. 3294-3296, Jul. 2020.
[40] X. Chen, W. Wu and B. Zhang, "Robust restoration method for active distribution networks," *IEEE Trans. Power Syst.*, vol. 31, no. 5, pp. 4005-4015, Sep. 2016.
[41] IEEE 123 node test feeder. [Online]. Available: https://cmte.ieee.org/pes-testfeeders/wp-content/uploads/sites/167/2017/08/feeder123.zip
[42] Cisco 1240 Connected Grid Router Hardware Installation Guide, Aug. 2019. [Online]. Available: https://www.cisco.com/c/en/us/td/docs/routers/connectedgrid/ cgr1000/hardware/ cgr1240/installation.pdf
[43] T-H. Yang and C-C. Tsai, "Using numerical weather model outputs to forecast wind gusts during typhoons," *Journal of Wind Engineering and Industrial Aerodynamics*, vol. 188, pp. 247-259, May. 2019.
[44] J. Li, Z. Song, X. Wang, Y. Wang and Y. Jia, "A novel offshore wind farm typhoon wind speed prediction model based on PSO–Bi-LSTM improved by VMD," *Energy*, vol. 251, p. 123848, Jul. 2022.
[45] Gurobi Optimization, LLC. 2021. [Online]. Available: https://www.gurobi.com